\documentstyle[emulateapj]{article}
\def\ltsima{$\; \buildrel < \over \sim \;$}
\def\simlt{\lower.5ex\hbox{\ltsima}}
\def\gtsima{$\; \buildrel > \over \sim \;$}
\def\simgt{\lower.5ex\hbox{\gtsima}}
\def\be{\begin{equation}}
\def\ee{\end{equation}}

\lefthead{Wang et al.}
\righthead{BAL QSO}
\begin{document}
\title{THE X-RAY ABSORBER IN BROAD ABSORPTION LINE QUASARS}
\author{T. G. WANG,\altaffilmark{1}, W. BRINKMANN,\altaffilmark{2} 
W. YUAN,\altaffilmark{3}, J.X. WANG,\altaffilmark{1} Y.Y.
ZHOU\altaffilmark{1}}
\altaffiltext{1}{Center for Astrophysics, University of Science and Technology
of China, Anhui, 230026, CHINA (email:twang@ustc.edu.cn)}
\altaffiltext{2}{Max-Planck-Institut f\"ur extraterrestrische Physik,
Giessenbachstrasse,D-85740 Garching, FRG}
\altaffiltext{3}{Space Astrophysics Group, National Space Development Agency of
Japan (NASDA), Tsukuba Space Center, Sengen-2-1-1, Tsukuba, Ibaraki 305 JAPAN
}
\authoremail{twang@ustc.edu.cn}

\begin{abstract}

Recent observations of Broad Absorption Line (BAL) quasars demonstrated 
that the 
soft X-ray emission of  these objects is extremely weak
and convincing evidence for very strong absorption by a high column 
density ($\sim 10^{23.5}$ cm$^{-2}$) was obtained for PG
1411+442, even though it is one of the few BAL QSOs strongly detected in soft
X-rays. This paper examines the ionization status and geometry of the
X-ray absorber by combining the properties of the UV lines with the X-ray
continuum absorption. We show that the gas has to have  large
column densities in ions of major UV absorption lines, such as CIV, NV, OVI
and Ne VIII, in order to have sufficient opacity around 0.2 to 0.35 keV. 
The UV absorption lines have to be saturated if the X-ray absorber
intersects the line of sight to the UV continuum emission region.
A uniformly covering UV and X-ray absorption model can be constructed for PG
1411+442  but in some other soft X-ray detected BAL QSOs, such as PG
1001+054, the observed line optical depth is much lower than expected from 
the X-ray absorbing material. We propose a scheme in which a substantial
fraction of the line of sight to the continuum source may be covered by either
an optically thick flow or clouds in a narrow velocity range, but in which the
total covering factor of either the whole flow or all clouds is close to unity.

The absorber can contribute significantly to the extremely highly ionized 
emission lines, such as O VI 1032/1037\AA~ and Ne VIII 770/780\AA~ if it 
covers a substantial fraction of solid angle and if the density is higher than 
$10^8$ cm$^{-3}$. However, it has very little impact on the medium and 
low ionization UV lines such as NV, CIV. 
The profiles of NeVIII and OVI lines may be indicators for the
kinematics of the X-ray absorber in QSOs. The observed Ne VIII line profiles
in QSOs suggests that the velocity of the gas projected
onto our line of sight is similar to that seen in the outflows 
of the UV BALs.

\end{abstract}
 
\section{Introduction}

Broad Absorption Lines (BALs) are prominent features in the UV spectra of
$\sim$ 10-15\% of optically selected quasars. The lines are formed in ionized
winds in which the speeds can range from near zero to more than 30,000
km~s$^{-1}$. The absorption features most commonly observed are Ly$\alpha$
$\lambda$ 1216, CIV $\lambda$1549, SiIV $\lambda$1397, and NV
$\lambda$1240. Low ionization broad absorption lines (LoBALs), such as
MgII $\lambda$2798 and Al III $\lambda$1857 are detected in about 15\% of
the objects. For two BAL QSOs, higher ionization lines, up through Ne VIII $\lambda$774,
Mg X $\lambda$615, Si XII $\lambda$499, are reported from  UV
observations with HST and HUT (Korista \& Arav 1997, Telfer et al. 1998).
Since most BAL QSOs are only observed by ground based optical telescopes which
can not access the spectral range of these very high ionization lines, the
true incidence of the very high ionization BALs is not known.

The observed frequency of BALQSOs among all quasars, combined with the assertion
that the absorbing material covers $\le$ 0.2  of the solid angle, 
leads to the conclusion that
every QSO has outflows which intercept only a fraction of the solid angle of the
central object (e.g., Hamann, Korista \& Morris 1993). This picture is supported by the similarity of UV continuum and
emission line properties of BAL and non-BAL QSOs (Weymann et al., 1991). The BAL region covers, 
at least partially,  the broad emission line region since the NV
absorption trough is usually deeper than the local continuum. The large
momentum of radiation absorbed  by resonance transitions is probably 
responsible 
for the acceleration of the gas. The signature of a kick-off effect at
velocities of about 6000 km s$^{-1}$ further supports this picture (Arav, Li \&
Begelman 1994).

Establishing the physical properties of the flow is a fundamental issue in
BAL studies. Earlier work was solely  based on a few prominent UV
absorption lines, usually CIV, Ly$\alpha$, NV, SiIV, and, for LoBAL, MgII and
AlIII, with the assumption that the absorbing material fully covers the
emission region and is not seriously saturated. These studies suggested an 
over-abundance of metals; C and N are, relative to H, enhanced by a
factor 3-5 and an even larger abundance for Si was deduced.
 More recently, based on the detection
of P~V absorption lines,  Haman (1998)  argued 
for a large column density instead of extremely high phosphorus abundance.  
From the UV spectra taken by the Hubble Space Telescope of high or moderate
redshift BAL QSOs, the line profiles of many  more lines, including 
lines from different ionization stages of the same element, were measured
for several  objects. These observations indicate that either the
absorbing material only covers partially  the line of sight to the continuum or
the scattered light fills in the absorption trough (e.g., Arav et al. 1999). The fact that the 
spectro-polarimetry
of BAL usually shows higher polarization in the BAL trough than for the
continuum and emission line region also favors the latter interpretation (e.g.,
Schmidt \& Hines 1999, and references therein).

X-ray spectra, even at their currently relatively low resolution, have the
advantage that  photo-electric  X-ray absorption 
does not have saturation problems and that it can be used to measure the total
absorption column density. As a result, the light from the
scattering medium or the leakage of a partially covered absorber is 
distinguishable from the transmitted light, thus allowing measurement of the
effective covering factor of the absorber.
 
BAL QSOs had not been targets for a systematic study in the X-ray
band until their extra-ordinary properties in the soft X-ray band were
discovered by ROSAT (Green et al. 1995, Green \& Mathur 1996), and they remain a rare class with extremely
weak X-ray  emission. In the soft X-ray band, there are only 
four reported  cases of
detections (PG 1411+442, PG 1001+054, PG 2112+059, SBS 1542+541
(Brinkmann et al. 1999, Brandt et al. 1999, Telfer et al.
1998)  showing   apparent broad 
band optical-to-X-ray spectral indices $ 1.85 \leq \alpha_{ox} \leq 
 2.25$, while only upper limits are available for the other objects. ROSAT
observations of the LoBAL QSO PG 1700+518 yield an even higher upper limit
of $\alpha_{ox} > 2.3$ (Wang et al. 1996), and the current X-ray
observations are not sensitive enough to establish how high these values 
really are.
 However, we noticed that a ROSAT detection is more likely
for weak BAL objects. Two ROSAT detections, Mrk 231 and
IRAS 07598+6508 (marginal), are probably associated with the
circumnuclear starburst instead of nuclear activity (Turner 1999).
The weak X-ray emission was interpreted to be due to strong absorption, instead
of an intrinsically X-ray weakness of the sources.
 This required a column densities of $10^{23-24}$ 
cm$^{-2}$ (Brinkmann et al. 1999, Wang et al. 1999, Gallagher et al. 1999). 

Therefore, a critical question arises  whether the UV and X-ray
absorber is the same material. What is the geometry of this
absorber? What is the ionization status?  In this
paper, we try to address these questions by examining the ionization status of 
the X-ray absorber in the BAL.

\section{Ionizing Continuum}

The shape of the ionizing continuum is crucial for the determination of the
ionization structure of the photo-ionized gas. For non-BAL QSOs, 
an average spectrum is derived by combining the soft X-ray spectrum at the 
low redshift end and the HST far-UV spectrum at high redshifts (Laor et al.
1997, Zheng et al. 1998).
However, we cannot construct a mean ionizing continuum for BAL QSOs in
this way as soft X-ray emission has not been detected from most BAL QSOs.
In addition, it
is believed that the few detections do not show the intrinsic 
spectrum  but only the
scattered/leaked component of the primary emission. The only reported
detection of the primary X-ray component is PG 1411+442 in the medium 
energy X-ray band by ASCA.

The simultaneous analysis of the ROSAT and ASCA data of PG 1411+442 was
presented by Brinkmann et al. (1999); a more extended treatment combining
the UV data from HST and the X-ray data from ROSAT and ASCA can be 
found in Wang et al. (1999). They found a primary X-ray component absorbed 
by $2.5\times 10^{23}$ cm$^{-2}$ and scattered flux  at about 3-5 per 
cent of the intrinsic emission. 
It is worthywhile to point out that although they
assumed the photon index of high energy component ($\Gamma_{high}$) to be 2.0 
because it ($2.2_{-0.7}^{+0.8}$ for 2.7 $\sigma$ error bar) was not well constrained by 
the fit, the absorption is well constrained to be larger than
1.5~$10^{23}$~cm$^{-2}$ (at 90\% confidence level) even when the 
$\Gamma_{high}$ is left as a free parameter.  

Figure 1 shows the Spectral Energy Distribution (SED) of PG 1411+442 from
the optical to the X-ray band. The infrared to optical data have been taken 
from Neugebauer et al. (1987). The UV spectrum from IUE between 2000-3000\AA~is
extracted from the IUE final archive and corrected for the Galactic reddening
E(B-V)=0.03 (N$_H^G$=1.4~ $10^{20}$ cm$^{-2}$). Reddening corrected HST
spectra were taken from Wang et al. (1999). In the X-ray band, the observed
ROSAT and ASCA spectra  are shown (model A of Wang et al. 1999). 
A composite radio-quiet QSO spectrum from the optical to the X-ray band
obtained  by combining
the results from Zheng et al. (1998) and Laor et al. (1997) is  plotted
for comparison.  The absorption corrected infrared to X-ray spectrum of
PG 1411+442 is clearly similar to that of average radio-quiet quasars. For the
above broken power-law model with absorption as a free parameter, the intrinsic 
1350\AA to 1 keV spectral index is 1.55$_{-0.33}^{+0.20}$ (only taking into
account the X-ray flux uncertainty).   

Figure 1 also shows the SED of two other BAL QSOs, PG 1001+054 and PG 2112+059,
detected by ROSAT. The ROSAT spectra were fitted with an
absorbed power law.  Absorption ($N_H=2.8_{-2.4}^{+6.3}~10^{20}$ $cm^{-2}$) 
in excess of the Galactic one is not needed for PG 1001+054; 
therefore, the column density is fixed at the Galactic value ($1.9~10^{20}$
$cm^{-2}$), and the fit yields a photon index $\Gamma=3.5_{-0.4}^{+0.6}$. 
However, the 
X-ray flux at 1~keV is lower by a factor of 50 compared to the mean 
radio-quiet QSOs. This is unlikely to be intrinsic since the HeII
$\lambda$1640 is fairly strong in the HST spectrum (Fig. 2) of PG 1001+442 
with an equivalent width larger than that in the composite HST QSO spectrum 
( Zheng et al. 1998),  
implying that the ionizing continuum seen by the BLR should be hard 
(Korista, Ferland \& Baldwin 1997). The strength of CIV line, found to 
correlate 
with ionizing continuum spectral index (e.g., Wang et al. 1998),  also 
indicates hard continuum spectrum. 
 In fact, the apparent broad band spectra of PG 1001+059 and 
PG 1411+442  from the infrared to soft X-rays are similar. Therefore, we
believe that the ROSAT spectrum is also the scattered component, as in PG
1411+442, and that the fraction of scattering is even lower in this object.  
The interpretation of scattering is consistent with the fact that the
ROSAT spectral index of PG 1001+054 fits well to the FWHM of H$\beta$ versus
spectral index correlation (see  Wang, Brinkmann, \& Bergeron 1996, 
Laor et al. 1997).

The situation for PG 2112+043 is slightly more complicated. Large absorption is
certainly present. By fixing the photon index to the average  value of 2.7 of 
the radio-quiet PG quasars a power law fit with free absorption yields an 
optical to X-ray
spectral index 2.0$\pm 0.2$ and an absorption column density
$N_H$=7$_{-4}^{+8}$~10$^{21}$~cm$^{-2}$, where errors are given at 90\% confidence
levels for one parameter. Clearly, the absorption corrected flux is still
lower by a factor of about 5-10 with respect to that of average QSOs (see also
Figure 1). It is, however, 
not clear whether  the ROSAT spectrum represents the transmitted
continuum or a combination of scattered plus transmitted component.
Complicated absorption models such as a partially covered high column plus
a uniformly covering low column absorber can reproduce the feature as well.

In summary, so far it appears that  the observed broad SEDs of BAL QSOs are
consistent with the intrinsic SED of typical non-BAL QSOs modified by absorption
in the soft X-ray band. For this reason, we will use the average  quasar
spectral energy distribution in our calculations. 
However, there is no guarantee that it represents the 
typical SED of BAL QSOs, since the ROSAT detected BAL QSOs only account for a
small fraction of all BAL QSOs and there are some indications that they are
special; for example, the strength of the  
CIV BAL in these objects is lower than the average value for BAL QSOs
(Weymann et al. 1991).  

\section{Physical Parameters: General Considerations}

In this section, we summarize the ranges of the physical parameters,
obtained  either 
directly or estimated from observations. 
For the estimates derived from the UV spectrum, we assume that the X-ray 
absorber fully covers the UV continuum source.

{\it Column density}: The column density for PG 1411+442  has been 
determined  to be $N_H=2.5~10^{23}$ cm$^{-2}$.
 Although the column densities for other  BAL QSOs have not  directly
been measured, they are likely similar or even larger. 
The fact that the X-ray emission from BAL QSOs is very weak, even in
the ASCA band, constrains the column densities 
to be higher than  a few times $10^{23}$~cm$^{-2}$
in most bright BAL QSOs if their intrinsic X-ray emission is 
comparable to that of the 
other radio-quiet QSOs (Brinkmann et al. 1999, Gallagher et al. 1999). 
Thus, a typical column density of $10^{23.5}$~cm$^{-2}$ seems 
to be a quite conservative estimate. 

{\it Density}: The density of the absorbing gas is virtually unknown.
 However,  for a photo-ionized gas the ionization structure is not 
sensitive to the density. 
The lack of broad [OIII] emission in QSOs suggests that the typical density
of an UV absorber is higher than $10^8$ cm$^{-3}$. This conclusion,
however, is somewhat ionization-dependent.  If the BAL region is highly
stratified and Oxygen is ionized beyond [OIII] as our calculations
for X-ray absorber  show (see sect. 4.1),
then there are no constraints for a lower limit to the gas density 
in the highly ionized zone.  Earlier reports of a possible detection
of absorption lines from excited ions OV* and CIII* 
(Pettini \& Boksenberg 1986, Korista et al. 1992) require 
an even higher density ($n_e\simeq 10^{11}$ cm$^{-3}$). 
These results were questioned by later high S/N
observation (Arav et al. 1999). A rather large range of densities might
co-exist in the BAL region and Arav et al.  proposed  dense flow tubes
embedded in a low density global flow to explain the ionization dependent
covering factor.
   
{\it Metal abundances:}  BAL quasars are claimed to have 
remarkably high abundances of heavy
elements relative to the solar values.  C, Si, N, P, and Fe are over-abundant
by 1-2 orders of magnitude. The abundances were derived from the optical BAL
depths together with photoionization models assuming ionization equilibrium.
The enhancement of metal elements can be reduced, but not eliminated, by using
complicated shapes for the ionizing continuum. The qualitative results
will not be altered by considering saturation of the absorption lines
since troughs in CIV and NV are much deeper than in Ly$\alpha$.  
The calculations presented below, however, are
based on cosmic abundances, since little is known about the abundances
of the X-ray absorber, which might be different from that of the UV absorbing gas. 
Since the X-ray opacity is mainly determined by heavy elements, the
column density mentioned above actually measures that of the heavy elements;
therefore, the results are correct as far as the metals
are concerned. 	The hydrogen column density is roughly inversely proportional 
to the measured metal abundance. 
The reason is that for lower abundances thicker material is needed to
produce the same X-ray absorption, and the X-ray column density is the 
equivalent hydrogen density with solar metal abundances. 

{\it Ionization parameter}: An upper limit on the ionization parameter
for the X-ray absorber can be obtained 
by requiring that the material is not too highly ionized  to be 
transparent for the low energy soft X-rays.  Brinkmann et al. (1999)
showed that a uniformly covering warm absorption model does not fit the
ROSAT+ASCA spectra of PG 1411+442, and they argued for a partially covering
absorption model or for an absorption plus scattering model.  However, both
Brinkmann et al.  and Wang et al. simply used a neutral absorber. We
show below that combining the ROSAT and ASCA spectrum can put a tight
upper limit on the ionization parameter of the absorber. 

Allowing the
absorbing medium to be  partially ionized, we use the ionized absorption
model {\em absori} in XSPEC to replace {\em wabs} for the model of an 
absorbed, broken power law plus a scattering component by electrons (see Wang
et al. 1999 for details). 
The break energy has been fixed at 1.0 keV and the
photon index of the hard X-rays at $\Gamma$=2.0. 
The ionizing continuum is assumed to be
a power law with a photon index 2.5 from the UV to hard X-rays.
This simplified continuum is not an exact representation 
for the ionizing continuum but it produces
similar results as Netzer's more complicated model for PG 1126-041 
(H. Netzer, private communication).
  We use an X-ray ionization parameter
U$_x$, which is defined as the dimensionless 
ratio of photon density above 0.1~keV and
the particle density (Netzer 1996), instead of $\xi=L/(nr^2)$, since 
it is much less dependent on the detailed UV to X-ray spectral
slope.  Figure 3 shows 68\%, 90\% and 99\%  contours for the ionization
parameter versus the column density for the absorbing material. The upper
limit on $U_x$ is mainly constrained by the photon flux at 
energies lower than 0.4~keV, and depends only weakly on the column density. 
At $N_H = 2~10^{23}$ cm$^{-2}$, one obtains $U_x < 0.11$ (at 90\% confidence
level, see Fig. 3).   
For the SED of average radio-quiet quasars, linearly extrapolated from the UV 
to X-ray energies, the 
hydrogen ionization parameter is $U_H = 35~U_x$.

If we assume that the intrinsic X-ray spectrum of PG 1001+054 is similar to an 
average radio-quiet QSO, the unabsorbed X-ray flux will be by a factor 30 
greater.  We fitted the ROSAT spectrum with a two component model, in which a 
primary power law 
continuum, 30 times larger than the observed one, is completely covered by an 
ionized absorber and an electron-scattered continuum component which is only 
absorbed by the Galactic absorption, similar to the model for PG 1411+442. 
Taking typical column density ($N_H= 10^{23.5}$~cm$^{-2}$) for the ionizing 
material, an upper limit (at 90\% confidence level) of 
 $U_x < 0.2 $ can be estimated. The exact value is dependent on the assumed
column density as well as assumption of the intrinsic continuum. 


A lower limit on the ionization parameter can be obtained by considering the
absorption by hydrogen. In the HII zone, the fraction of neutral hydrogen is
approximately $n(H^0)/n(H^+) \simeq 10^{-5.3}/U_H$ for a gas photo-ionized 
by a typical AGN continuum (Netzer et al. 1990). The ratio would be lower 
than that given above when the ionization parameter is high, so that 
collisional ionization becomes
significant. Since the Lyman break is not severe for most HI BAL QSOs 
(Korista et al. 1992, Arav et al. 1999) 
 this puts  an upper limit on the total column density of
neutral hydrogen on the line of sight to the UV continuum emission region
to be less than $10^{17.3}$~cm$^{-2}$ ($\tau_{912}\simeq 1$).
For typical column densities this implies  a lower limit of $U_H = 5$ 
if the X-ray absorber fully covers the UV continuum emission region. 
A tighter constraint on the neutral hydrogen column density can be derived by 
using the Ly$\alpha$ line absorption but required a knowledge of the
distribution of HI column density in velocity space.

\section{Predicted Absorption and Emission Lines}

\subsection{Absorption Lines}

In order to qualitatively estimate the ionization status of the heavy 
elements, we first consider the source of opacity in the 0.2 to 0.4 keV 
soft X-ray band. For a weakly ionized gas helium is a major contributor
to the opacity. 
However, if the hydrogen Lyman break is not 
severe, helium atoms will produce negligible absorption in 0.2-0.4 keV 
band. The fraction of He$^+$ can be roughly estimated 
from  $He^{++}/He^{+} 
\simeq 10^{3.2} U_H$ for the typical ionizing AGN continuum (Netzer et al.
 1990). This gives 
$H^0/He^+ \simeq 0.1$ for gas dominated by $He^{++}$. Since 
the photo-electric absorption cross section decreases with photon energy 
as $E^{-3}$, in order to produce an optical depth of order $\tau = 1$ at 0.2 keV, 
an optical depth $\tau(54 eV)\simeq 51$ is required!
This implies an optical depth at  the 
Lyman limit of the order of 9. Therefore it appears unlikely that helium 
contributes significantly to the opacity at 0.2 keV without producing 
a serious Lyman break.
Thus the 0.2 to 0.4 keV opacity must be produced by heavy elements.   
     
Figure 3 shows the ionization potentials (IPs) of the valence electrons
for the ions of the four most abundant elements C, O, N, and Ne. 
The IPs for the
first K-shell electrons are 392, 552, 739 and 1196 eV for C, N, O and
Ne, respectively.  From this plot, it is clear that, in order to have a 
significant opacity below 0.35 keV, a substantial fraction of the elements
O and Ne must keep at least one L-shell electron, i.e.,  they are not
ionized beyond
O$^{5+}$ and Ne$^{7+}$. Note that O$^{5+}$ and Ne$^{+7}$ are also ions 
which produce major resonant UV absorption lines. The fact that resonant 
line absorption has a much larger cross
section than photo-electric absorption (by a factor of order 10$^4$)
suggests that the optical depth of the line absorption is much larger than
the absorption edge in the soft X-ray band if  OVI and NeVIII are
responsible for the X-ray absorption. 
The  large absorption in the soft
X-rays at energies as low as 0.3 keV requires that  OVI and NeVIII absorption
lines in the X-ray absorbing gas must be optically very thick.  This
general result is not strongly dependent on the specific SED or on the relative
metal abundances of O and Ne.  A detailed ionization calculation 
(see below)  shows  that  
in photo-ionization equilibrium a substantial fraction of CIV and NV
should also exist in this case.
 
For the calculations a 
 plane-parallel geometry is adopted and the ionizing continuum  for
average radio-quiet quasars  illuminates one side.
  A typical column density of $10^{23.5}$ cm$^{-2}$ is assumed.
  The ionization equilibrium is determined by using the
photoionization code CLOUDY 94.01 (Ferland 1999). Figure 4 shows the 
ionization structure of the 
elements carbon, nitrogen, oxygen and neon for an ionization parameters U$_H$=10. 
In this plot, C$^{3+}$, N$^{+4}$, O$^{+5}$, Ne$^{+6}$, and Ne$^{+7}$
coexist over a very large region. The opacity at energies between 0.1keV and 
0.35~keV are mainly due to the ions O$^{+5}$, Ne$^{+7}$ and Ne$^{+6}$. 

The transmitted continua are plotted in  Fig. 5 for U$_H$ = 8,10, and 16.
A substantial Lyman edge is seen only in the first case, 
but a substantial HeII edge at 
54.4 eV is present in all cases. This is because in a gas ionized by a
typical AGN-continuum, $He^{++}/He^+\simeq 10^{3.2}U_H$. This means that 
the column density of $He^+$  is larger than that of H$^0$ by a factor of
nearly 10 in a fully
ionized gas. It can be seen that only when U$_H \le $10
for the given column density, a large fraction of the soft X-rays between
0.2-0.3 keV can be absorbed. This corresponds to  $U_X\simeq 0.29$,
which is roughly consistent with that  derived from the X-ray spectral fits
for PG 1411+442. The soft X-ray opacity between 0.1 to 0.3 keV is mainly 
caused by OVI, NeVII, and NeVIII ions, whose abundances are very sensitive 
to U$_H$, as can be seen in the figure.  

Table 1 shows the predicted column densities for U$_H$ = 10 and U$_H$ = 8. As one can
see from the table, the absorber can certainly predict a large column density 
in ions producing the major UV absorption lines within the allowed range of 
ionization parameters. 
At U$_H$=10, the model predicts  CIV and NV column densities close to the total 
column densities estimated from strong BALs, and an OVI column density much 
larger than previously estimated based on the UV absorption line. An even 
larger column density is produced for NeVIII lines. Further, the column density
 of neutral hydrogen is an order of magnitude higher than required by Ly$\alpha$
absorption.  
 Since the column densities of these ions of CIV, NV, OVI and NeVII increase 
rapidly with decreasing ionization parameters and U$_H$ = 10 is an upper limit 
imposed by requiring strong absorption in the 0.2-0.35 keV band, the above column
densities represent only lower limits.    

Since both, Ly$\alpha$ absorption and the Lyman edge, are sensitive to 
the exact value of the ionization parameter and the metal abundances, the 
lower limit for the ionization parameter can be dramatically relaxed 
for much higher metal abundances or a very 
steep ionizing continuum.  Since the fraction of neutral
hydrogen is only proportional to $U_H^{-1}$, the reduction of the fraction of
hydrogen by an order of magnitude, while keeping the X-ray opacity constant 
would require the ratio $U_H$ to $U_x$ to be ten times larger, implying 
a very steep UV to X-ray spectrum $\alpha_{uvx}\simeq 2.3$. This 
seems very unlikely in view of the results of 1411+442, where 1350\AA~ to ~1 
keV spectral index $\alpha_{uvx}= 1.55_{-0.33}^{+0.20}$ when corrected for 
the intrinsic absorption (see section 2). 
If the metal abundances are one order of magnitude higher, the hydrogen
column can be lower by a similar factor, because the absorption in the
X-ray band is mainly due to metal elements.

In order to see how the predicted column densities change with the assumed 
hydrogen column density, we calculated similar models for $N_H=10^{22.5}$ 
 and 10$^{23}$ cm$^{-2}$. The density and ionizing continuum were the same as 
used in the previous calculations. In order to limit the transmitted 0.2-0.3 
keV flux to less than 30\% of the primary, the ionization parameters ($U_H$) 
had to be less than 1.4 and 3.6 for $N_H=10^{22.5}$ and 10$^{23}$ cm$^{-2}$,
respectively. For these $U_H$'s, the predicted column densities of
 C$^{3+}$, N$^{4+}$, 
O$^{5+}$, Ne$^{6+}$, and Ne$^{7+}$ are similar to those of the
 $N_H=10^{23.5}$ cm$^{-2}$ case (see
Table 1), as expected. We did not calculate the column density for the
$N_H=10^{24}$ cm$^{-2}$ model since the escape probability approximation used in  
Cloudy may result in an incorrect  ionization structure (Dumont, 
Abrassart \& Collin 2000). But the argument presented at the beginning of this
section suggests that the ion column densities should be similar.   
  
\begin{deluxetable}{llccc}
\tablecaption{Predicted column densities for different BAL ions}
\tablehead{
\colhead{log $N_H$} & \multicolumn{2}{c}{23.5} & 23 & 22.5 \\
                     \cline{2-3} \\
\colhead{U$_H$} &\colhead{10} & \colhead{8} & \colhead{3.6$^{a}$} & \colhead{1.4$^{a}$} 
}
\startdata
CIV    & 2.4$\times$10$^{16}$  & 8$\times 10^{16}$ & 3.2$\times 10^{16}$  & 2.8$\times 10^{16}$ \nl
NV     & 3.4$\times$10$^{16}$  & 9$\times 10^{16}$ & 4.7$\times 10^{16}$ &  4.8$\times 10^{16}$ \nl
OVI    & 1.2$\times$10$^{18}$  & 4$\times 10^{18}$ & 1.8$\times$10$^{18}$ &
1.9$\times$10$^{18}$ \nl
NeVII  & 5.2$\times$10$^{17}$  & 4$\times 10^{18}$ & 7.4$\times$10$^{17}$  &7.5$\times$10$^{17}$ \nl
NeVIII & 3.7$\times$10$^{18}$  & 6$\times 10^{18}$ & 2.3$\times$10$^{18}$  & 1.4$\times$10$^{18}$\nl
HI     & 3.0$\times$10$^{16}$  & 4$\times 10^{16}$ & 3.4$\times$10$^{16}$ &
4.0$\times$10$^{16}$ \nl
\enddata
\tablenotetext{a}{The photo-ionization parameters for a single zone
constrainted to transmit about 30\% of the intrinsic 0.2-0.3 keV soft X-ray
flux.}
\end{deluxetable}

Although our calculations are only carried out for a one-component model,
more complicated models, such as multiple-zone models, should 
produce strong CIV, NV, OVI absorption lines as well. 
 In order to isolate the zone which 
produces the absorption lines, we considered a two-zone model. The
X-rays at energies at $>$0.35 keV are absorbed by a high column, highly
ionized absorber, in which the C, N, O atoms are ionized beyond CIV, NV,
OVI, and at lower energies by a low column density, low ionization
material. Since the high ionization zone contributes little to the
X-ray absorption below 0.35 keV, the opacity below 0.35 keV has to be  
dominated by the low ionization material. The main contribution to the
opacity between 0.2-0.4 keV is from  N, O, and Ne at 
moderately high ionization parameters, and from He at low ionization 
parameters, when $He^+$ is the dominating species. In the former case, an 
argument similar to the one given at the beginning of this section suggests 
large absorbing column densities in CIV, NV, OVI, NeVII and NeVIII.   
In the latter case, a large optical depth at the Lyman limit is predicted, 
and strong absorption by lines from low ionization species such
as OIII, CIII, NIV,  etc.  seems inevitable. 

\subsection{Emission Lines}

Absorption lines can only trace the outflowing material along the line of sight, 
while information on the global distribution of the gas has to be deduced 
by other means. Emission lines represent one such tool. 

From the ionization structures given by the above calculations it is clear that the 
X-ray absorbing gas cannot contribute much to the emission lines from low 
ionization species. The line emission, unlike the ionization structure, depends
critically on the particle density as well as on the EUV and soft X-ray 
spectral slope. 
 For a density of 10$^9$ cm$^{-3}$ and U$_H$ = 10,
the model predicts EWs of 31\AA~ for OVI $\lambda$1032\AA~ and 21\AA~ 
for the NeVIII $\lambda$770/780\AA~ emission line,
for a  covering factor of 1.0. These values are a factor of two and  
three higher than the average EWs for these lines measured by HST (Hamann et
al. 1998). The line EWs increase to 77\AA~  for OVI  and 39\AA~ for Ne VIII 
for U$_H$ = 8. 
The absorber can make a significant contribution to the emission of NeVIII 
and OVI unless it covers only a small fraction of the solid angle or the 
density of the gas is small. In fact, Hamann et al. (1998) discussed already
the possibility of NeVIII emitting gas associated with the warm absorber in 
the X-ray band. 
Hamann et al. (1998) found that the NeVIII lines are significantly broader than
CIV in PKS 0355-483 and PG 1522+101. From their figures, the line widths extend
over the velocity range seen in typical BAL QSOs. 

\section{COMPARISON WITH OBSERVATIONS}

PG 1411+442:  Since the residual fluxes in the CIV and NV absorption
troughs are similar to the fraction of the unabsorbed X-ray flux, Wang et
al. (1999) argued that they are severely saturated at the bottom, but the 
observed residuals are the leaked light.
This makes any estimate of the total column densities of
CIV and NV very difficult and only lower limits can be derived. The
minimum column densities required to reproduce the observed CIV and NV
absorption features are of the order of $10^{16}$ cm$^{-2}$ for C$^{3+}$ and 
N$^{4+}$, and thus, are consistent with the expected lower limits to the 
column densities of these ions from the X-ray absorber estimated in section 
4.  
 The SiIV absorption line appears not to be saturated in 
this object; a column density of 2~10$^{14}$ cm$^{-2}$ (EW$\simeq$ 6\AA) is 
derived. Substantial SiIV can be present only in the zone where the carbon 
 is dominated by CIV or lower stages. The ionization parameter ($U_H\simeq 5$) 
is lower than the upper limit set by the soft X-ray absorption in the 
last section in order to reproduce the SiIV absorption line for the
column of 2~10$^{23}$ cm$^{-2}$ derived from the X-ray absorption.
This results in a higher column density (10$^{16-17}$ cm$^{-2}$) of HI 
for cosmic abundances, and higher CIV and NV column densities.
The box-shaped profile with substantial residual flux in the $Ly\alpha$ 
absorption line trough suggest that the absorption line is saturated, but 
the covering factor is smaller than 1 ($\sim$ 0.8). Thus, the 
observed HI column density can be consistent with the expectation from the 
X-ray absorber. The lower covering factor (0.8) may be due to the 
partially covering of the broad emission line region by the absorbing gas 
as part of Lyman emission line may be produced in a region more extended than 
CIV emission region. Thus, in PG 1411+442, the observed UV and X-ray data are 
consistent with a uniformly covered absorber model. Notice that the situation 
should be similar for other BAL QSOs with saturated absorption line 
troughs. 

PG 1001+054: If the intrinsic broad band spectrum of this object is
similar to other radio quiet AGN, i.e., $\alpha_{ox}=1.65$, the
detected soft X-ray component can not be the transmitted one as it is
rather steep.  Most likely, we have a similar scenario as in 
PG 1411+442.  The fraction of scattering is comparable, a few percent. 
Since the residual flux in the CIV absorption line
in this object is more than 30\% of the continuum, even at the  
deepest part of the trough, the observed profile implies either that
 the optically thick absorbing material is leaky and covers
less than 70\% of the sky or that the gas is not optically thick 
to the continuum photons. 
With the latter assumption, the total column density in CIV ions
is estimated to be $\simeq$2 $10^{15}$ cm$^{-2}$ (EW$\simeq$13\AA). 
 Since the  ROSAT spectrum is very steep, no leaking component is seen,
not  even in the hard  band of the ROSAT PSPC, and  therefore an 
absorption  column 
density of at least a few times 10$^{22}$ cm$^{-2}$ is required for X-ray
absorber.
Similarly, to keep sufficient opacity in the low energy soft X-ray band 
requires
that a substantial fraction of O and Ne atoms must be at ionization 
stages equal to or lower than OVI and NeVIII.
If the continuum is not too different from the average quasar SED, the X-ray
absorber would produce column densities of a few 
10$^{16}$~cm$^{-2}$ for 
CIV and NV (see table 1). This number is much larger than that inferred 
from the UV 
absorption lines. The discrepancy can hardly be explained by source variability 
as one needs a factor of 10 brighter X-ray emission in order to match 30\% 
residuals in the absorption line trough, considering the fact that 
statistically no BAL QSOs show strong X-ray emission.  Although we cannot 
rule out that an agreement can be achieved by using a completely different 
ionizing 
continuum, that
of Mathews \& Ferland (1987) with a bump in the  EUV still predicts 
a column density in CIV $> 10^{16}$~cm$^{-2}$ if the transmitted flux between
0.2-0.3 keV is less than 30\%. For a power law spectrum for the EUV 
( from 13.6 eV to 0.2 keV ) ionizing continuum, the
spectral index $\alpha$ ($f_\nu\propto \nu^{-\alpha}$) is required to be 
larger than 2.2 in order to match the observed CIV column density. This 
seems unlikely in view of our discussion presented in section 2.
 
Therefore, we prefer a picture in which an optically thick absorber partially 
covers the UV continuum source. There are two different type pictures for 
the partial covering. 
In the first case, the X-ray absorbing material
only partially covers the UV continuum emission region, but fully covers
the X-ray continuum region. This means that the UV
emission region is significantly larger than the X-ray emission region. 
However, it requires that the absorbing material is close to the central 
source. This seems unlikely as the NV absorption line profile indicates 
that the BAL gas is located at a distance at least as large as the emission 
line region. 

An alternative picture is that the line absorption at a given velocity is 
saturated 
but the absorbing material at the velocity only partially covers the line 
of sight to 
the continuum source, resulting in an apparently low column density in the
absorption line.
In order to explain the nearly complete absorption in the soft X-ray band, we
require that the global covering factor of the absorber, via integrating the
covering factor at different velocities, is  equal or very close to unity. This
can be realized by allowing the material at different velocities to intersect  
 the line of sight to different parts of the continuum. The partial covering 
at a given velocity is either due a  patchy absorber with covering factor 
of less than one or due to continuous outflow with 
 cross section smaller than the continuum source. 
One consequence of this kind of model is 
that the covering factor at any specific velocity is dependent on the
angle between the line of sight and the iso-velocity surface. In the clouds
model, larger viewing angle produces larger covering factor since more clouds on the
iso-velocity surface enter the line sight to the continuum source.   
 The opposite behavior is predicted for the small cross-section outflow
model since the covering factor at a specific velocity in this case is the 
projected area of the iso-velocity surface onto the line of sight to the 
UV continuum divided by the area of the UV emission region, which decreases 
with the viewing angle. With this picture, the different covering factor
of PG 1411+442 and  PG 1001+054 might be just due to an aspect effect,  
instead of intrinsically different covering factor of this absorber.         
Future X-ray observation of the photo-electric iron 
K edges should be able to reveal such velocity dependent partially covering 
features.

It is worthwhile to point out that the situation for the ROSAT undetected low
redshift bright QSOs with shallow BAL troughs is similar to that of
 PG 1001+054. However, this is hard to confirm for high redshift BAL QSOs
because the 0.2-0.35 keV band is shifted out of the ROSAT windows. Due to strong
Galactic absorption below 0.1 keV, future missions more sensitive at soft 
X-rays
might also have difficulties to settle this question for the BAL QSOs beyond
redshifts of 1.5.     

SBS 1542+541:
This is one of the highly ionized UV absorption line BAL QSO. The CIV and
NV absorption lines are weak, but absorption lines from OVI or higher
ionization stage ions are strong.
The optical to X-ray spectral slope was found to be 
$\alpha_{ox} = 1.86$ using ROSAT data (Telfer et al. 1998), 
which is significantly smaller  than the value 
found for other BAL QSOs. The S/N ratio of the ROSAT data is too low to 
allow detailed spectral fits. 
Since the ionization of BAL gas is much higher than conventionally observed in
other BAL QSOs, it is likely that the ionization parameter is also higher in
this object, and thus, it  appears
possible that a directly transmitted continuum is detected by ROSAT.
Given the redshift of z = 2.361 for this object, ROSAT actually measured the 
continuum in the 0.4-8.0 keV band in the rest frame of the QSO. 
The transmission of X-rays at medium energies is not very sensitive to the 
exact value of the  ionization parameter and is only a function of the 
total column density,
provided that the ionization parameter is not too high. 
Telfer et al. (1998) proposed an ionization dependent covering factor model 
to explain the strengths of UV absorption lines for cosmic abundances. The 
covering factor of low ionization ionic species such as HI is much smaller than 
those of high ionization species (up to 80\%). But they did not make any 
specific statement about the global covering factor of the gas. 
This model is consistent with the ROSAT data for this source. However, 
if we want to extend this model to other BAL QSOs, such as PG 1001+043, we 
have to require that the total covering of the absorber is close to unity 
in order to explain their weakness in the ROSAT band as we proposed for 
PG 1001+054.

\section{Conclusion}

We show that the opacity of soft X-rays at energies between 0.2-0.35 keV is 
dominated by L-shell photo-electric absorption of oxygen and neon in the
X-ray absorber of BAL QSOs. A stringent limit on the ionization parameter 
can be set by requiring presence of sufficient column densities for the ions 
responsible for the absorption in the soft X-ray band, namely, OVI, NeVII and 
 NeVIII. For typical equivalent hydrogen column density of 
$10^{23.5}$~cm$^{-2}$, the ionization parameter is required: $U_x< (0.2-0.3)$, 
or $U_H<10$ for the typical spectral energy distribution of
radio-quiet QSOs.  We find that the X-ray absorber has to have very large 
column densities in OVI, NeVII and NeVIII of the order of 10$^{17-18}$ 
cm$^{-2}$ in order to have sufficient opacity in the soft X-ray band. Since 
these column densities are derived from the observed optical depth in the soft
X-ray band, they do not strongly depend on the specific ionization continuum 
and the detail metal abundance.   
For reasonable ionizing continuum, it will also produce large column densities in CIV, NV and HI 
of order a few 10$^{16}$~cm$^{-2}$.

These results are compared with the observed X-ray and UV spectra for several
BAL QSOs with available soft X-ray data.  In PG 1411+442, a uniformly covering 
absorption model can explain the observed UV and X-ray absorption. However, 
in PG 1001+054, the UV absorption lines predicted from the X-ray absorption 
are much stronger than the observed ones, if the X-ray absorption material 
uniformly covers the UV continuum source.  
The discrepancy can be solved by a locally partial covering model in which 
the thick flow or clouds at a specific velocity covers only a certain 
fraction of the line sight to the continuum, but the total covering factor 
of all clouds or the flow is close to 1. As  
the covering factor of the absorber at any specific projected velocity is
strongly viewing angle dependent, this model has enough flexibility to produce 
the different BAL behaviors.   This model  might 
also apply to those BAL QSOs for which ROSAT yielded stringent 
upper limits on the X-ray flux but where the UV continuum is only 
partially absorbed. 

Our work shows that, assuming the X-ray and UV absorbers being the same, 
self-consistent absorption models can be constructed to explain the X-ray and
UV observations of the BAL QSOs treated herein; we thus suggest it as a 
promising case in BAL objects. However, future observations with a better 
sample and better quality data are need to draw firm conclusion on this,
considering following caveats of the analysis presented in this paper: 
(1) The QSOs herein, and especially PG 1411+442, may not be representative 
of BAL QSOs. (2) the assumption that the intrinsic SEDs are like those 
observed in other non-BALQSOs used in deriving the observed CIV and OVI column
density,  

While the absorber can contribute significantly to the extremely highly ionized
emission lines, such as O VI 1032/1037\AA~ and Ne VIII 770/780\AA~ if it
covers a substantial fraction of solid angle and if the density is higher than
$10^8$ cm$^{-3}$, it has very little impact on the medium and
low ionization UV lines such as NV, CIV and Ly$\alpha$.
The profiles of NeVIII and OVI lines may be indicators for the
kinematics of the X-ray absorber in QSOs. The observed Ne VIII line profiles
in QSOs suggests that the velocity of the gas projected
onto our line of sight is similar to that seen in the outflows
of the UV BALs.

\acknowledgements
We thank S. Komossa for critical reading of this manuscript and an anonymous
referee for helpful comments.
TW acknowledges the financial support from Chinese NSF (19925313) and from the 
Chinese Science and Technology Ministry.

\newpage
\begin{figure}
\figurenum{1}
\epsscale{1.00}
\plotfiddle{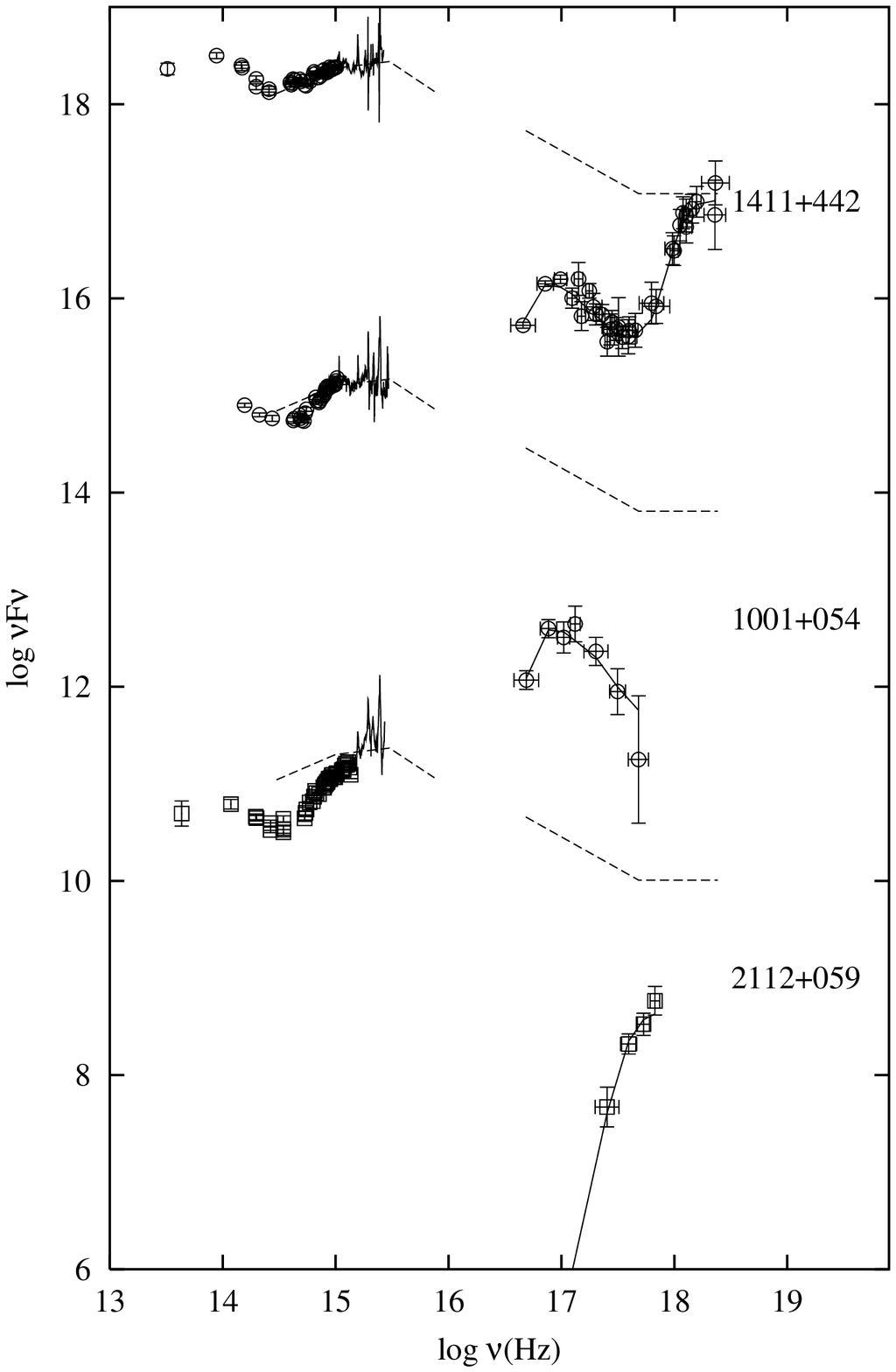}{13.5cm}{  0}{65}{65}{-200}{-70}
\figcaption[fig1.ps]{The Spectral Energy Distribution of three BAL QSOs
detected with ROSAT. The optical and UV spectra have been corrected for 
Galactic reddening. The dashed line shows the average SED for radio-quiet 
quasars  from  a composite of HST and the low redshift ROSAT PG quasar
sample (Zheng et al. 1998). Spectra have been shifted vertically arbitrarily for clarity.  
\label{fig-1}}
\end{figure}

\begin{figure}
\figurenum{2}
\epsscale{1.00}
\plotfiddle{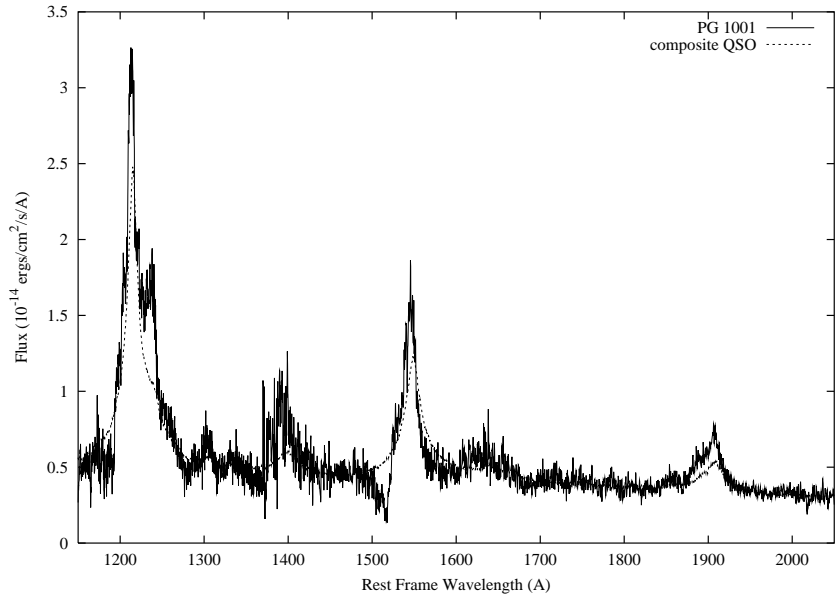}{7.cm}{270}{45}{45}{-200}{250}
\figcaption[fig2.ps]{A comparison of the UV spectrum of PG 1001+054 (solid
line) with 
the HST composite QSO spectrum (dotted line) showing strong HeII $\lambda$1640 and other
high ionization lines.
\label{fig-2}}
\end{figure}

\begin{figure}
\figurenum{3}
\epsscale{1.00}
\plotfiddle{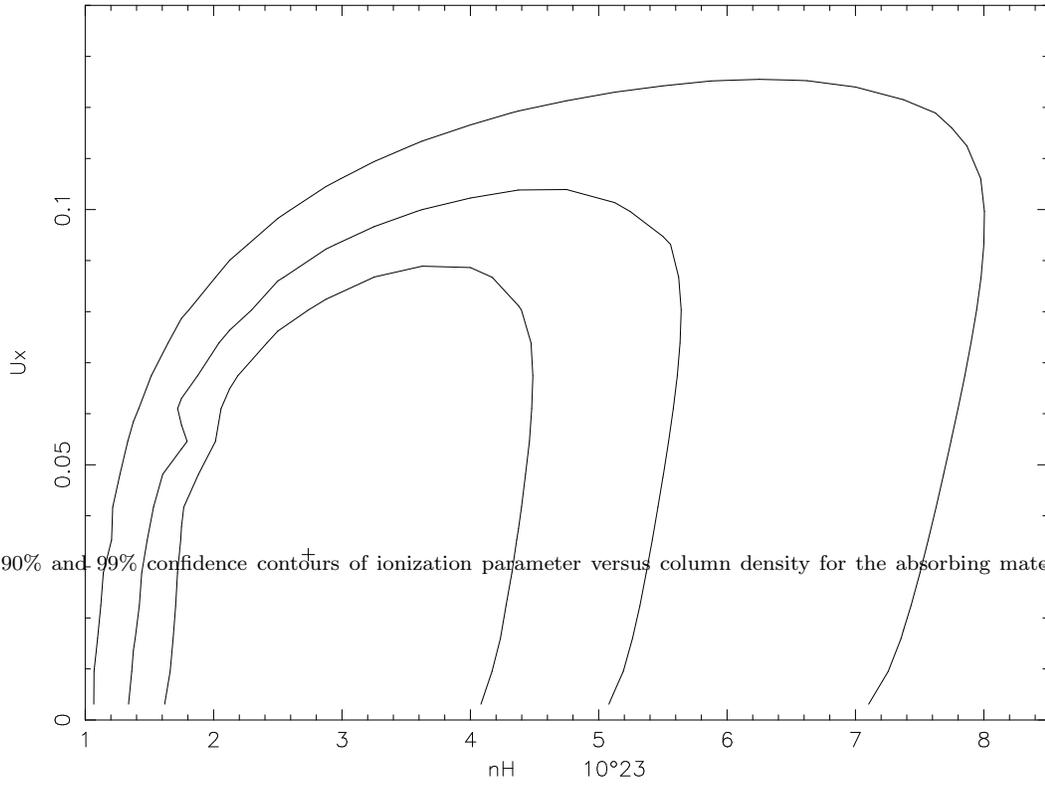}{7.cm}{270}{60}{60}{-200}{250}
\figcaption[fig3.ps]{The 68\%, 90\% and 99\% confidence contours of ionization
parameter versus column density for the absorbing material in PG 1411+442. 
\label{fig-3}}
\end{figure}

\begin{figure}
\figurenum{4}
\epsscale{1.00}
\plotfiddle{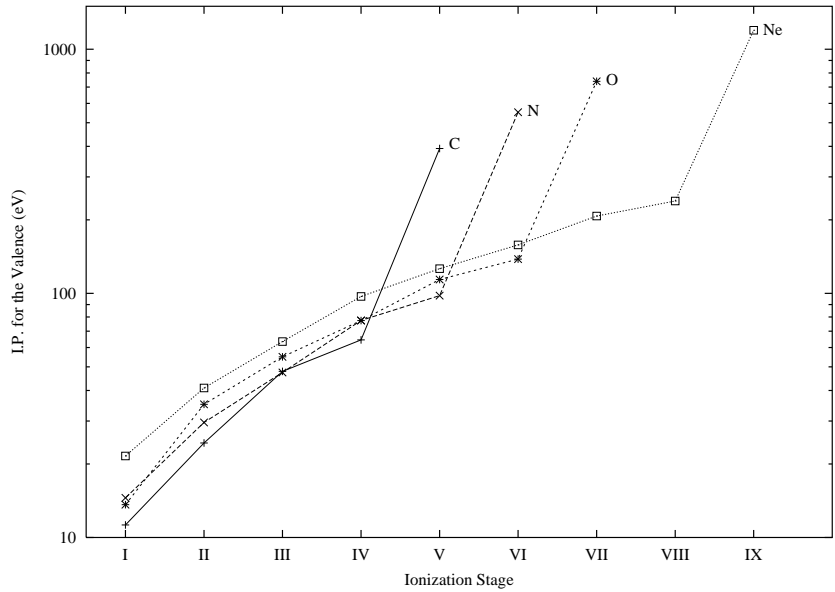}{7.cm}{270}{45}{45}{-200}{250}
\figcaption[fig4.ps]{Plot of ionization potential versus ionization stage for
valence electrons for elements C, N, O and Ne. Notice 
that the ionizing potentials for K-shell electrons are larger than 0.39 keV 
for these elements.  
\label{fig-4}}
\end{figure}

\begin{figure}
\figurenum{5}
\epsscale{1.00}
\plotfiddle{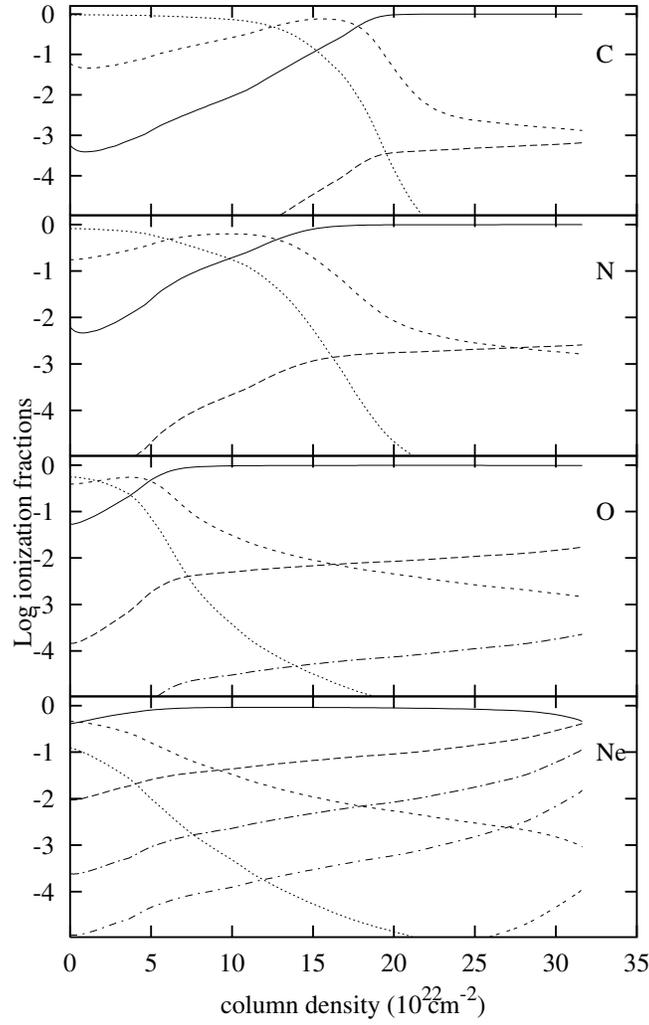}{14.cm}{  0}{70}{70}{-200}{-80}
\figcaption[fig5.ps]{The ionization structure of the elements C, N, O, and Ne 
for an ionization parameter 10 and a column density 10$^{23.5}$ cm$^{-2}$.
The curves show the logarithm of fraction of ion species, starting with a fully
stripped ion on the left, with decreasing 
ionization stages toward the right. 
\label{fig-5}} 
\end{figure}

\begin{figure}
\figurenum{6}
\epsscale{1.00}
\plotfiddle{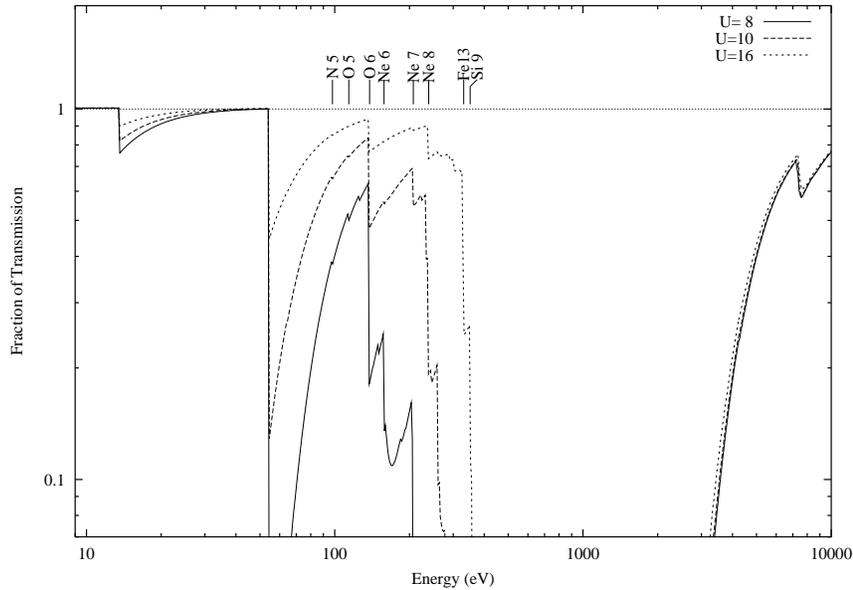}{7.cm}{270}{45}{45}{-200}{250}
\figcaption[fig6.ps]{The transmission rate of the ionization continuum in the 
10eV to 10 keV energy range for U$_H$~=~8, 10, and 16 corrected for  
Thomson scattering. 
\label{fig-6}}
\end{figure}
\end{document}